\documentclass[aps,pra,twocolumn,superscriptaddress,longbibliography]{revtex4-1}

\usepackage{amsmath}
\usepackage{amssymb}
\usepackage{amsfonts}
\usepackage{graphicx}
\usepackage[T1]{fontenc}
\usepackage{color}
\usepackage{textcomp}

\begin{document}

\title{Influence of excited state decay and dephasing on phonon quantum state preparation}

\author{Thilo~Hahn}
\affiliation{Institute of Solid State Theory, University of M\"{u}nster, 48149 M\"{u}nster, Germany}

\author{Daniel~Groll}
\affiliation{Institute of Solid State Theory, University of M\"{u}nster, 48149 M\"{u}nster, Germany}

\author{Tilmann~Kuhn}
\affiliation{Institute of Solid State Theory, University of M\"{u}nster, 48149 M\"{u}nster, Germany}

\author{Daniel~Wigger}
\email[]{d.wigger@wwu.de}
\affiliation{Institute of Solid State Theory, University of M\"{u}nster, 48149 M\"{u}nster, Germany}

\begin{abstract}
The coupling between single-photon emitters and phonons opens many possibilities to store and transmit quantum properties. In this paper we apply the independent boson model to describe the coupling between an optically driven two-level system and a discrete phonon mode. Tailored optical driving allows not only to generate coherent phonon states, but also to generate coherent superpositions in the form of Schr\"odinger cat states in the phonon system. We analyze the influence of decay and dephasing of the two-level system on these phonon preparation protocols. We find that the decay transforms the coherent phonon state into a circular distribution in phase space. Although the dephasing between two exciting laser pulses leads to a reduction of the interference ability in the phonon system, the decay conserves it during the transition into the ground state. This allows to store the phonon quantum state properties in the ground state of the single-photon emitter.
\end{abstract}

\date{\today}

\maketitle

\section{Introduction}

The link between nanophotonics and phononics has recently been gaining more and more attention. The reason is that many promising studies have shown that phonons -- besides giving rise to dephasing of optical transitions in nanosystems \cite{muljarov2004deph,machnikowski2006chan} -- offer a robust and controllable way to act on nanosystems. It has been demonstrated that externally excited surface acoustic waves \cite{naber2006surf,gell2008modu,couto2009phot,schulein2015four} or coherent bulk acoustic waves \cite{bruggemann2012lase} can be used to tune the transition energies of single-photon emitters, such as excitons in quantum dots or defect centers in diamond \cite{golter2016opto}. For example, based on this modulation of the transition energy, the output of a semiconductor microcavity laser can be either enhanced or completely switched off \cite{czerniuk2014lasi,wigger2017syst}.

The coupling between electrons and phonons can also be utilized in the opposite way. As a consequence of rapid optical excitations of such single-photon emitters, phonons are generated \cite{mahan2013many}. Whereas the excitation of acoustic phonons leads to the emission of phonon wave packets \cite{krummheuer2005pure,hohenester2007quan,wigger2014ener}, longitudinal optical (LO) phonons are characterized by a negligible dispersion and, therefore, cannot leave the region of creation. In Ref.~\cite{reiter2011gene}, it was shown that a single ultrafast optical excitation may lead to the generation of coherent LO phonon states whereas two-pulse sequences can be used to generate Schr\"odinger cat states. Considering the LO phonon energies, which are in the range of tens of meV for typical III-V semiconductors~\cite{grundmann1995inas} or even at a few hundreds of meV for hexagonal boron nitride~\cite{wigger2019phon}, the phonon-related time scales are typically much faster than characteristic radiative decay or dephasing times of the corresponding single-photon emitters. Therefore excited state decay and dephasing are usually neglected in this context.

However, LO phonons are not the only discrete vibration mode that can be coupled to single-photon emitters. In the field of optomechanics, the eigenmodes of micrometer-sized solid-state resonators are investigated \cite{kippenberg2008cavi,aspelmeyer2014cavi}. Compared to LO phonon modes where the atomic masses determine the characteristic frequencies, the large resonators have huge masses resulting in very small frequencies \cite{unterreithmeier2009univ,o2010quan,teufel2011side,munsch2017reso}. The coupling between such resonators and single-photon emitters has already been studied theoretically \cite{wilson2004lase,wilson2007theo} and experimentally \cite{lee2016strai,munsch2017reso}. Because of the low oscillator frequencies, here the decay time scales of the single emitters become comparable with the phonon-related time scales. Hence, when considering phonon state preparation protocols by optical excitation of quantum dot excitons or defect centers interacting with large resonator systems, spontaneous decay, e.g., by photon emission, and dephasing have to be taken into account. In this paper, we systematically discuss this influence. We will show that, although excited state decay and dephasing only act on the electronic system, depending on the excitation conditions, they may strongly influence the quantum state of the coupled phonons. This influence also opens up new possibilities to generate specific phonon quantum states.

\section{Theory}

\subsection{Model}
We consider a two-level system (TLS), representing the single-photon emitter, coupled to a single discrete phonon mode. Since the energy of the phonon mode is typically much lower than the transition energy of the TLS and phonons, therefore, do not lead to transitions between the ground and the excited states, we describe the optically driven coupled emitter-phonon system by the standard single-mode independent boson Hamiltonian,
\begin{eqnarray}
H &=& \hbar \Omega \left| x\right>\left< x\right|
- \big[{\bf M\cdot E}(t) \left| x\right>\left< g\right| + {\bf M^*\cdot E^*}(t) \left| g\right>\left< x\right| \big] \notag\\
&&+ \hbar \omega_{\rm ph} b^\dagger b 
+ \hbar g \big( b + b^\dagger \big) \left| x\right>\left< x\right| \ ,
\label{eq:hamiltonian}
\end{eqnarray}
where $\left|g\right>$ and $\left|x\right>$ denote the ground and excited states of the TLS, respectively, and $\hbar\Omega$ is the transition energy. The bosonic operators $b$ and $b^\dagger$ act on a single-phonon mode with energy $\hbar\omega_{\rm ph}$ and corresponding oscillation period $t_{\rm ph}=2\pi/\omega_{\rm ph}$. The optical transitions, treated in terms of the usual rotating-wave and dipole approximations, are mediated by the dipole matrix element ${\bf M}$ and the optical field ${\bf E}(t)$. The phonon coupling to the TLS is treated via the pure dephasing mechanism with the coupling strength $g$, taken to be real.

In addition we include excited state decay (xd), e.g., by means of radiative decay due to the coupling to the photon vacuum, and pure dephasing (pd), e.g., due to the influence of a fluctuating environment, which are both assumed to be Markovian processes described by two phenomenological Lindblad dissipators~\cite{breuer2002theory},
\begin{subequations}
\begin{eqnarray}
\mathcal{D}_i(\rho) 
&=& \eta_i \left[ \left(A_i^{ } \rho A_i^\dagger\right) - \frac12 \left\{A_i^\dagger A_i^{ },\rho \right\} \right] 
\end{eqnarray}
with
\begin{eqnarray}
A_{\rm xd} &=& \left| g\right>\left< x\right|\ ,\quad \eta_{\rm xd} = \Gamma\ , \\
A_{\rm pd} &=& \left| x\right>\left< x\right|\ ,\quad \eta_{\rm pd} = 2\tilde{\beta}\ .
\end{eqnarray}
\label{eq:lindblad}
\end{subequations}
Here, $\Gamma$ is the excited state decay rate labeled as xd, and $\tilde{\beta}$ is the pure dephasing rate labeled as pd.

\subsection{Equations of motion} 
Based on the Hamiltonian \eqref{eq:hamiltonian} and the Lindblad dissipators \eqref{eq:lindblad}, the equation of motion for the density matrix of the coupled electron-phonon system reads
\begin{equation}
\frac{\rm d}{{\rm d}t}\rho = \frac{1}{i\hbar} \left[H,\rho \right] + \mathcal{D}_{\rm xd}(\rho)+ \mathcal{D}_{\rm pd}(\rho)\ . \label{eq:EOM}
\end{equation}
As has been discussed in Ref.~\cite{axt1999cohe} for the single-mode case and in Ref.~\cite{vagov2002elec} for the multi-mode case, the quantum state of the coupled electron-phonon system can be completely described in terms of three generating functions for phonon-assisted density matrices defined according to
\begin{subequations}
\begin{eqnarray}
Y(-\alpha^\ast,\alpha, t) &=& \big< \left| g\right>\left< x\right| \exp (-\alpha^* b^\dagger) \exp (\alpha b)\big>\ , \\
C(-\alpha^\ast,\alpha, t) &=& \big< \left| x\right>\left< x\right| \exp (-\alpha^* b^\dagger) \exp (\alpha b)\big>\ , \\
F(-\alpha^\ast,\alpha, t) &=& \big< \exp (-\alpha^* b^\dagger) \exp (\alpha b)\big>\ .
\end{eqnarray}
\end{subequations}
Based on the general equation of motion \eqref{eq:EOM}, the equations of motion for the generating functions can be derived leading to
\begin{subequations}
\begin{eqnarray}
i\hbar\dot{Y} &=&
\hbar\big[ \Omega 
+ \omega_{\rm ph}(\alpha \partial_\alpha - \alpha^\ast \partial_{\alpha^\ast}) \notag\\
&&+ g(\alpha + \partial_\alpha -\partial_{\alpha^\ast}) \big] Y \notag\\
&+& {\bf M}\cdot {\bf E}(2C-F) - i\hbar\beta Y \ , \label{eq:EOM_Y}\\
i\hbar\dot{C} &=& 
\hbar\left[\omega_{\rm ph}(\alpha\partial_\alpha - \alpha^\ast\partial_{\alpha^\ast}) 
+ g(\alpha + \alpha^\ast)\right] C \notag\\
&+& {\bf M}^* \cdot {\bf E}^* Y - {\bf M}\cdot {\bf E} Y^T - i\hbar\Gamma C \ , \label{eq:EOM_C}\\
i\hbar\dot{F} &=& \hbar \omega_\text{ph} (\alpha\partial_\alpha - \alpha^\ast \partial_{\alpha^\ast})F + \hbar g(\alpha + \alpha^\ast)C \ , \label{eq:EOM_F} 
\end{eqnarray}\label{eq:EOM_tot}\end{subequations}
\noindent where $\beta = \tilde{\beta}+\Gamma/2$ and $Y^T(-\alpha^\ast,\alpha,t) = Y^\ast(\alpha^\ast,-\alpha,t)$. Following Ref.~\cite{axt1999cohe}, by introducing $Y(\alpha,t)=Y(-\alpha^\ast,\alpha,t)$ (and corresponding definitions for $C$ and $F$), the dimensionless phonon coupling strength $\gamma=g/\omega_{\rm ph}$ and the polaron-shifted transition energy $\overline{\Omega}=\Omega-\gamma^2\omega_{\rm ph}$ and using the transformations,
\begin{subequations}
\begin{eqnarray}
\overline{Y}(\alpha, t) &=& 
\exp\left( i\overline{\Omega} t + \gamma \alpha e^{i \omega_\text{ph} t} \right) 
Y\left(\alpha e^{i\omega_\text{ph} t} - \gamma, t\right)\,, \quad \\
\overline{C}(\alpha, t) &=& 
\exp \left[ i 2\gamma \text{Im}(\alpha e^{i\omega_\text{ph}t})\right] 
C\left(\alpha e^{i\omega_\text{ph} t},t\right)\,,\\
\overline{F}(\alpha, t) &=& 
F\left(\alpha e^{i\omega_\text{ph} t}, t\right)\,,\\
\overline{E}(t) &=& 
\frac 1\hbar {\bf M\cdot E}(t)e^{i\overline{\Omega}t}\,,
\end{eqnarray}
\end{subequations}
the partial derivatives with respect to $\alpha$ and $\alpha^\ast$ can be eliminated. The resulting equations of motion for the case without excited state  decay and dephasing can be found in Refs.~\cite{axt1999cohe,wigger2016quan}. Including decay and dephasing is straightforward. For arbitrary shapes of the driving electric field, they can be solved numerically. 

In Ref.~\cite{vagov2002elec}, it was shown that in the limit of excitation by an arbitrary sequence of ultrafast optical pulses the equations of motion for $\overline{Y}$, $\overline{C}$, and $\overline{F}$ can be solved analytically. From a physical point of view, the ultrafast limit is reached if the pulse duration is much shorter than any phonon-related time scale, here, in particular, much shorter than the phonon oscillation period $t_{\rm ph}$. In this case, the phonon influence on the dynamics during the excitation is negligible, and the pulse sequence can be mathematically well approximated by a series of $\delta$ pulses at times $t_j$, 
\begin{equation}
\overline{E}(t) = \sum_j \frac{\theta_j}{2} e^{i\overline{\Omega}t_j+i\varphi_j}\delta(t-t_j)
= \sum_j \frac{\theta_j}{2} e^{i\phi_j}\delta(t-t_j)\ ,
\end{equation}
with $\theta_j$ being the pulse area, $\varphi_j$ being the carrier-envelope phase, and $\phi_j$ is the total phase of the $j$th pulse arriving at $t=t_j$. 

Following the derivation in Ref.~\cite{vagov2002elec} but additionally including the decay rate $\Gamma$ for the excited state occupation and the total dephasing rate $\beta=\Gamma/2 + \tilde{\beta}$ for the interband coherence, we can split the time evolution into recursion relations that: (i) link the transformed generating functions at $t_{j}^-$ immediately before pulse $j$ with the ones at $t_{j}^+$ directly after this pulse and (ii) connect these functions at $t_{j}^+$ with the ones immediately before the subsequent pulse at $t_{j+1}^-$.

Between the pulses: (i) the coherence function $\overline{Y}$ decays due to the dephasing constant $\beta$, (ii) the excited state occupation function $\overline{C}$ decays due to the decay constant $\Gamma$, and (iii) the phonon function $\overline{F}$ is modified due to the excited state occupation. Denoting, e.g., the coherence function just before and after the pulse $j$ by $\overline{Y}_{j}^{\pm}=\overline{Y}(t_{j}^{\pm})$ and correspondingly for the other functions, the connection between the functions just after pulse $j$ and immediately before pulse $j+1$ read
\begin{subequations}
\begin{eqnarray}
\overline{Y}_{j+1}^{-} &=& e^{-\beta (t_{j+1}-t_{j})}\overline{Y}_{j}^{+} \,,\\
\overline{C}_{j+1}^{-} &=& e^{-\Gamma (t_{j+1}-t_{j})}\overline{C}_{j}^{+} \,,\\
\overline{F}_{j+1}^{-} &=& \overline{F}_{j}^{+} + \overline{C}_{j}^{+} \\
&\times & \int\limits_{t_{j}}^{t_{j+1}} e^{-\Gamma (t-t_{j})} \frac{\partial}{\partial t} \exp\left(-i 2 \gamma{\rm Im} \left[ \alpha e^{i \omega_{\rm ph} t}\right] \right)\, {\rm d}t\,. \notag
\end{eqnarray}\label{eq:free}\end{subequations}

Due to the ultrashort pulse limit, the recursion relations across the pulses are not influenced by decay and dephasing. Therefore, they agree with those derived in Ref.~\cite{vagov2002elec}, reduced to the single-mode case. For completeness, these relations, in the notation of the present paper, are summarized in the Appendix in Eq.~\eqref{eq:pulse}. Note that, in particular, the phonon function $\overline{F}$ is not changed during the pulses.

\begin{figure}
\includegraphics[width =\columnwidth]{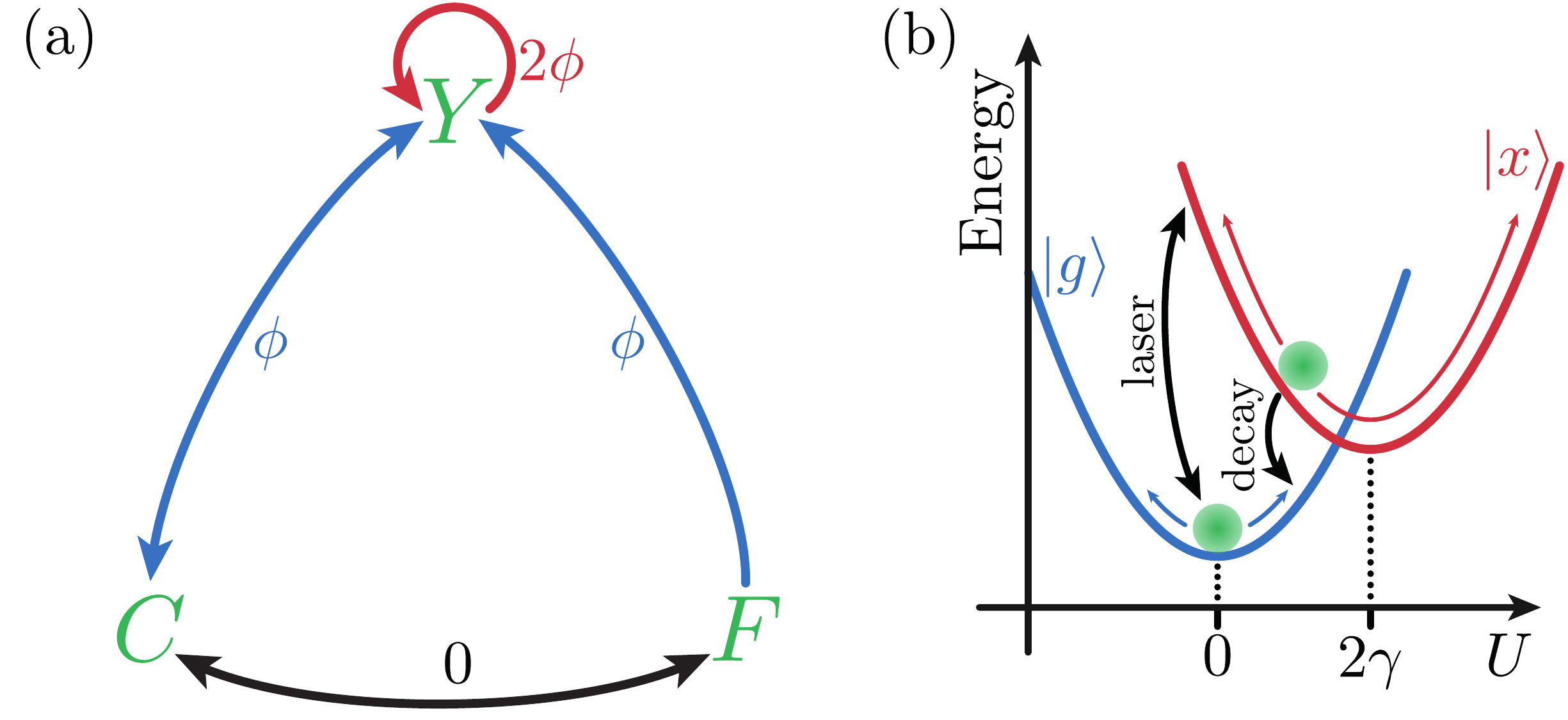}
\caption{Schematics of the system and its temporal evolution. (a) Mutual influence of the generating functions caused by a laser pulse with phase $\phi$ [Eq.~\eqref{eq:pulse}] and during the evolution between the pulses [Eq.~\eqref{eq:free}]. The labels at the arrows indicate the phase that is transferred in the respective coupling. (b) Illustration of the phonon potentials associated with the ground state $\left|g\right>$ (blue) and the excited state $\left|x\right>$ (red) of the TLS. The phonon state is depicted by the green dots. Laser excitation and decay induce transitions between the electronic states.}
\label{Fig:1}
\end{figure}

Figure~\ref{Fig:1}(a) schematically shows the interplay among the three generating functions $Y$, $C$, and $F$ as described by Eqs.~\eqref{eq:free} and \eqref{eq:pulse}. The arrows reflect the influence of a laser pulse with phase $\phi$ and the couplings induced by the phonons. The labels at the arrows indicate the order of $\phi$ entering the coupling. We see, for example, that the phonon function $F$ is only driven by the excited state occupation $C$, whereas it influences both, the occupation $C$ and the coherence $Y$.

\subsection{Wigner representation}
We are primarily interested in the properties of the phonon system. As its central quantities, we define the displacement $u$ and momentum $p$ via the quadratures,
\begin{equation}
u = (b + b^\dagger)\ ,\qquad p = \frac{1}{i} \left(b - b^\dagger\right)\ ,
\end{equation}
with their eigenfunctions $u\left| U\right> = U\left| U\right>$ and $p\left|\Pi\right> = \Pi\left|\Pi\right>$. In Fig.~\ref{Fig:1}(b), it is schematically shown that the phonon system is split into two subsystems, one associated with each state of the TLS. The phonons attributed to the excited state $\left|x\right>$ have an equilibrium displacement, which is shifted with respect to the ground state $\left|g\right>$ by $2\gamma$ \cite{reiter2011gene}. The laser pulse excitation mediates transitions between the two systems, whereas the decay of the excited state acts only in one way from $\left|x\right>$ to $\left|g\right>$.

A useful tool to study the quantum state of the phonon system is the Wigner function, which is defined via the phonon density matrix as
\begin{equation}
W(U, \Pi) = \frac{1}{4\pi} \int \left< U + \frac{X}{2}\right| \rho_{\rm ph} \left| U-\frac{X}{2}\right> e^{-i X\Pi/2}\,{\rm d}X\ ,
\end{equation}
with $\rho_{\rm ph}={\rm Tr}_{\rm TLS}(\rho)$, where we take the trace with respect to the TLS. It is especially handy as it can directly be calculated from the generating function $F$ via the Fourier transform,
\begin{equation}
W(U,\Pi) =  \frac 1{4\pi^2} \int e^{-|\alpha |^2/2} F(\alpha) e^{i\left[{\rm Re}(\alpha)\Pi+{\rm Im}(\alpha) U\right]}\,{\rm d}^2\alpha \ . \label{eq:W}
\end{equation}
The Wigner function is a quasiprobability distribution in the phase space spanned by $U$ and $\Pi$. Although it can take negative values the calculation of expectation values, of all operators which can be expressed as (symmetrized) functions of $u$ and $p$ is performed, as in the case of common probability density functions. For example, it is 
\begin{subequations}
\begin{eqnarray}
\left<u\right> &=& \iint U\, W(U,\Pi)\,{\rm d}U{\rm d}\Pi\quad {\rm or}\\
\left<u^2\right> &=& \iint U^2\, W(U,\Pi)\,{\rm d}U{\rm d}\Pi\ .
\end{eqnarray}\end{subequations}

By performing the same transform as in Eq.~\eqref{eq:W} but with $C(\alpha)$ instead of $F(\alpha)$, we isolate the part of the phonon state that is associated with the excited state $W_{\rm x}(U,\Pi)$. With this, we can also identify the part of the ground state $W_{\rm g}$ from
\begin{equation}\label{eq:Wsum}
W = W_{\rm g}+W_{\rm x}\ .
\end{equation}

\section{Results}
In the following sections, we will discuss the influence of excited state decay and dephasing on the properties of the generated phonon state. Before addressing specific excitation scenarios, let us briefly come back to the schematic in Fig.~\ref{Fig:1}(a). There it is seen that the phonon function $F$, while influencing both other functions $C$ and $Y$, is only influenced by the occupation function $C$. This tells us that, in the case of excitation by a single ultrashort pulse, the phonon dynamics cannot be influenced by pure dephasing processes. In contrast, in the case of excitation by multiple pulses, pure dephasing occurring between the pulses influences the light-induced TLS dynamics of later pulses and, thus, indirectly also the phonon properties.

To clearly identify the role of excited state decay and pure dephasing processes on the dynamical phonon state, we start by considering a single laser pulse excitation and afterwards move to two-pulse excitations where we will first investigate the influence of each process separately and finally look at their combined effect. 

\subsection{Single-pulse excitation}\label{sec:decay}
As it has been discussed in Ref.~\cite{reiter2011gene} without considering excited state decay and dephasing, a single-pulse excitation of the TLS with a pulse area of $\theta=\pi$ instantaneously shifts the equilibrium position of the entire phonon system. The initial phonon vacuum state, thus, becomes a coherent phonon state that oscillates around a displaced equilibrium position in phase space at $(U,\Pi)=(2\gamma,0)$.
Considering an excitation at $t=0$ with pulse area $\theta$ but taking into account a nonvanishing decay rate $\Gamma$, the characteristic phonon function $F(t)$ after the pulse, i.e., for $t\geq0$, reads
\begin{eqnarray}
F(\alpha,t) &=& 1-i2\gamma\omega_{\rm ph}\sin^2\left(\frac{\theta}{2}\right)\int\limits_0^t e^{-\Gamma t^\prime} {\rm Re}\left[\alpha e^{i\omega_{\rm ph} (t^\prime - t)}\right] \notag \\
&\times& \exp\left\{ i2\gamma {\rm Im}\left[\alpha e^{-i\omega_{\rm ph} t}(1- e^{i\omega_{\rm ph} t^\prime})\right]\right\}\, {\rm d}t^\prime\, . 
\label{eq:F_1}
\end{eqnarray}
Inserting this in Eq.~\eqref{eq:W} leads to the corresponding Wigner function of the phonon system.

\begin{figure}
\includegraphics[width = 1.0\columnwidth]{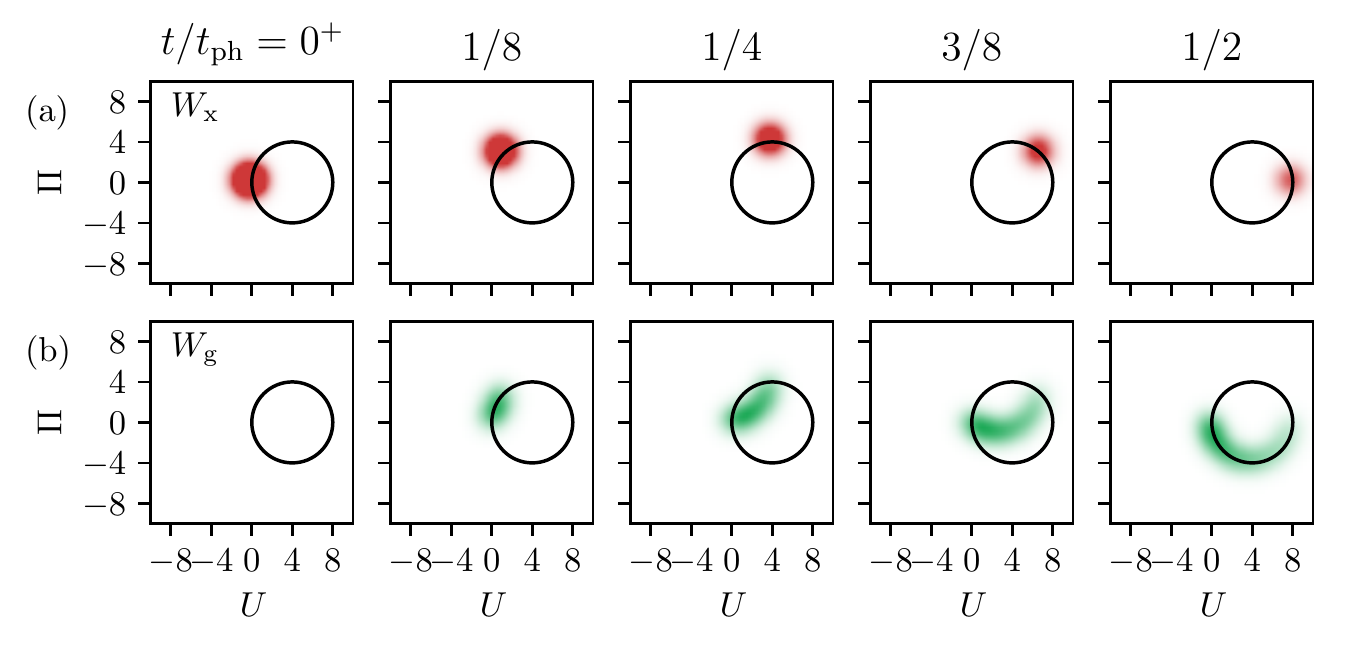}
\caption{Partition of the phonon Wigner function into the (a) excited state and (b) ground state subspace. The corresponding subspace Wigner functions are shown at five different times after excitation by a $\pi$ pulse. The system parameters are $\gamma=2$, $\Gamma=0.5\,\omega_{\rm ph}$.}
\label{Fig:2}
\end{figure}

As schematically shown in Fig.~\ref{Fig:1}(b), the phonon state is partially attributed to the ground state $\left|g\right>$ of the TLS and the rest to the excited state $\left|x\right>$. The pulse area $\theta$ in Eq.~\eqref{eq:F_1} determines to which amount the system is brought into the excited state and, therefore, which fraction of the phonon state switches into the shifted potential. A nonvanishing decay rate of the excited state makes the phonons go back to the unshifted potential associated with the ground state of the TLS as depicted in Fig.~\ref{Fig:1}(b). 

Let us start by considering the excitation by a pulse with $\theta=\pi$ in a system with an electron-phonon coupling constant $\gamma=2$ and an excited state decay constant $\Gamma=0.5\,\omega_{\rm ph}$. Snapshots of the Wigner functions describing the dynamics of the phonons corresponding to the excited state $W_{\rm x}$ and to the ground state $W_{\rm g}$ are plotted for five different times in Figs.~\ref{Fig:2}(a) and \ref{Fig:2}(b), respectively. A $\pi$ pulse completely inverts the system, therefore, immediately after the pulse ($t=0^{+}$), the phonon system is completely attributed to the excited state. The initial state is a coherent state in this subspace, which rotates around the shifted equilibrium position (black circle), much like in the case without decay. The excited state decay leads to a gradual damping of the phonon state in the excited state subspace. The ground state contribution $W_{\rm g}$ in Fig.~\ref{Fig:2}(b) is more involved. Starting with an empty phase space at $t=0^{+}$, the Wigner distribution builds up during the dynamics. The excited state successively decays into the ground state, which adds contributions to $W_{\rm g}$ at the point in phase space where the coherent state in the excited state in Fig.~\ref{Fig:2}(a) is located. Subsequently, these contributions continue to rotate, but since the equilibrium position for $W_{\rm g}$ is the origin $(U,\Pi)=(0,0)$, this rotation occurs on circles around the origin. Interestingly, at the last considered time $t/t_{\rm ph}=1/2$, they all lie on the lower half of the circle reflecting the trajectory of $W_{\rm x}$. Indeed, it is clearly seen that at this time we obtain a Wigner function $W_{\rm g}$ distributed along the circular shape of this trajectory.

\begin{figure}
\includegraphics[width = 1.0\columnwidth]{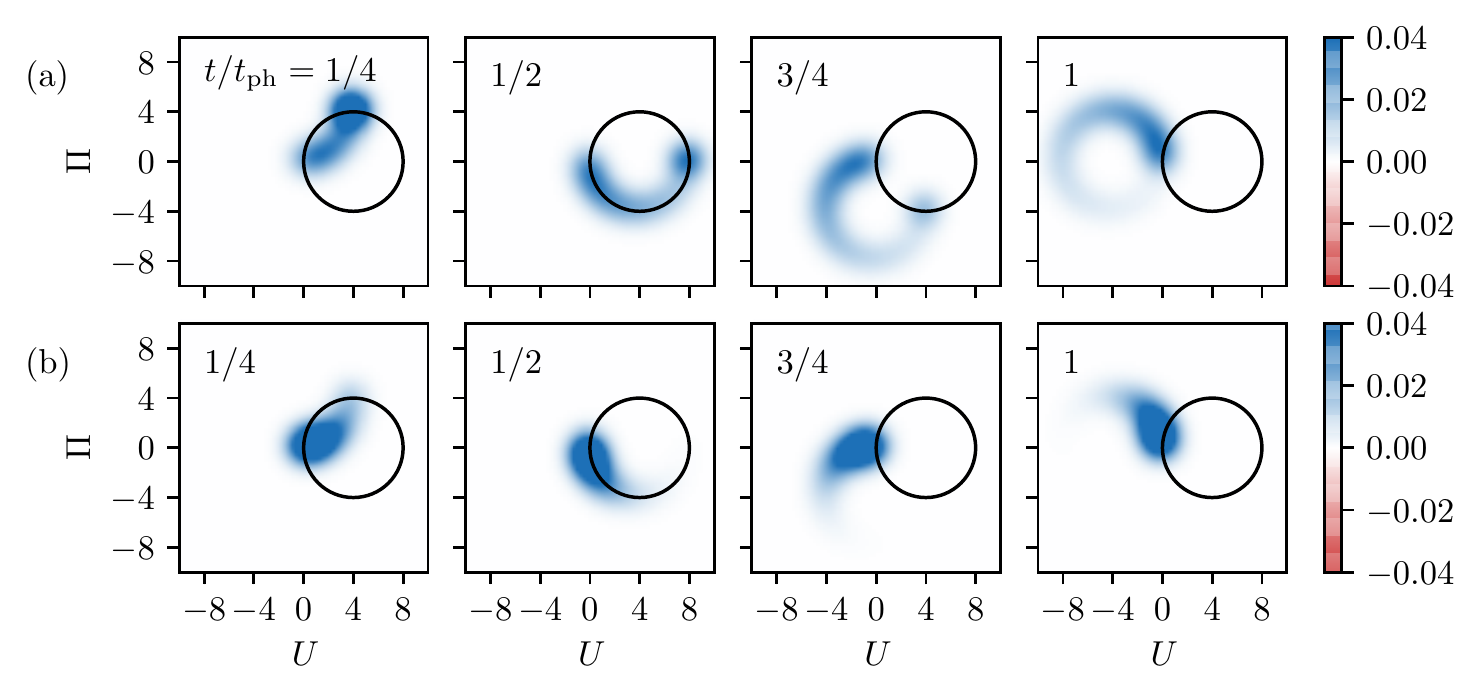}
\caption{Phonon Wigner function at four different times after excitation by a single $\pi$ pulse for an electron-phonon coupling $\gamma=2$ and decay constants (a) $\Gamma=0.5\,\omega_{\rm ph}$ and (b) $\Gamma=2\,\omega_{\rm ph}$.}
\label{Fig:3}
\end{figure}

The total phonon Wigner function is the sum of the ground and the excited state contributions [Eq.~\eqref{eq:Wsum}]. Snapshots of this function are shown at four different times for the same case as discussed above in Fig.~\ref{Fig:3}(a). For short times after the laser pulse excitation ($t/t_{\rm ph}=1/4$), a large fraction of the distribution moves on the marked circle and keeps its Gaussian form. However, the influence of the decay already manifests in a curved tail connecting the main Gaussian part with the origin of the phase space. The reason for this shape is simply the sum of the two contributions $W_{\rm g}$ and $W_{\rm x}$ discussed before. After half a phonon period ($t/t_{\rm ph}=1/2$), the fraction of the Gaussian on the marked circle has become weaker whereas the curved tail has gained weight. This process goes on until the excited state has entirely decayed into the ground state. After a full phonon period at $t=t_{\rm ph}$ in Fig.~\ref{Fig:3}(a), the Wigner function is distributed along a circle of the same size as the marked one, but shifted in the opposite $U$ direction. At this time, the excited state occupation is $\left<\left|x\right>\left<x\right|\right>\approx 0.1\%$ meaning that a large fraction of the phonon state already oscillates around $(0,0)$. When the TLS has entirely decayed into the ground state, the Wigner function will rotate  stable in shape around the origin of the phase space. 

When we increase the decay rate to $\Gamma=2\,\omega_{\rm ph}$, the phonon state's dynamics expressed in terms of the Wigner function is depicted in Fig.~\ref{Fig:3}(b). Without any decay, the Wigner function would move stable in shape on the black circle around the shifted origin at $(U,\Pi)=(2\gamma,0)=(4,0)$. However, we find that, already in the first shown time step ($t/t_{\rm ph}=1/4$), the largest part of the distribution has left the marked circle of the coherent state. The reason is that the TLS has already decayed by almost $80\%$ into the ground state. The overall shape of the Wigner function is similar to the one in Fig.~\ref{Fig:3}(a) but with a shorter tail, i.e., it has a larger weight near the origin of the phase space. 

\begin{figure}
\includegraphics[width = 1.0\columnwidth]{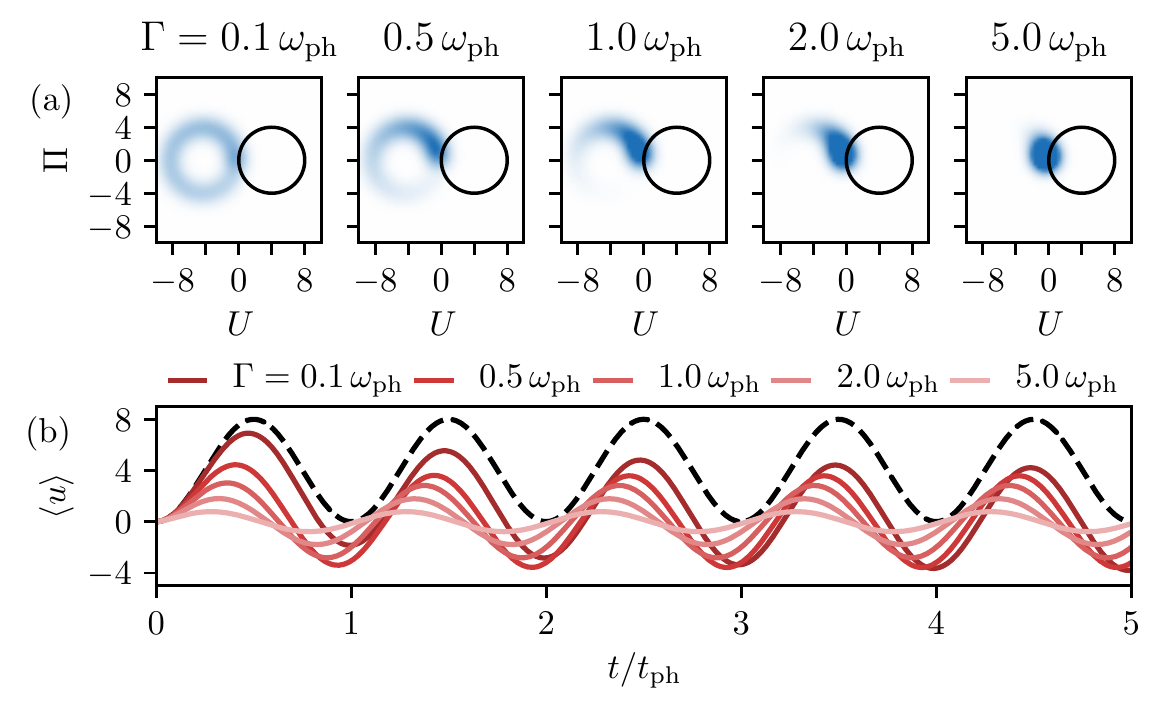}
\caption{(a) Phonon Wigner function after excitation by a single $\pi$ pulse at the time $t/t_{\rm ph}=10$, i.e., after the excited state occupation has completely decayed for an electron-phonon coupling $\gamma=2$ and five different values of the decay constant $\Gamma$. (b) Expectation value of the displacement $\left<u\right>$ as a function of time after the laser pulse for the same values  of $\Gamma$. The dashed line shows the expectation value for the case without decay (i.e., $\Gamma=0$).}
\label{Fig:4}
\end{figure}

Figure~\ref{Fig:4}(a) summarizes the influence of the excited state decay on the phonon state by showing the Wigner function for different values of the decay constant at $t/t_{\rm ph}=10$ when the TLS has completely decayed.  The Wigner function is located on a circle which rotates around the origin. At integer values $t/t_{\rm ph}=n$ as shown here, the circle is a mirror image of the trajectory of the coherent state without decay (black circle), whereas at half-integer values $t/t_{\rm ph}=n+\frac{1}{2}$, it coincides with this trajectory. For decay constants much smaller than the phonon oscillation period, the Wigner function is essentially homogeneously distributed along the circle, whereas for increasing decay constants, it becomes more and more concentrated on a part of the circle. Finally, if the decay is much faster than the oscillation period, the excited state occupation has already decayed before the phonons could react on the excitation. Therefore, in this case the phonons essentially remain in the initial vacuum state.

The Wigner function reflects the full quantum state of the phonon system. The resulting temporal evolution of the mean displacement is depicted in Fig.~\ref{Fig:4}(b) for the same parameters as in part (a) of the figure. The dashed line displays the dynamics of the mean displacement for vanishing excited state decay. As discussed above, the phonons form a coherent state with a displacement oscillating around $2\gamma$, i.e., in the present case around $\left< u \right> = 4$. Including excited state decay, the equilibrium position of the phonons returns to zero. Therefore, at long times, all solid curves in Fig.~\ref{Fig:4}(b) oscillate around zero. For weak damping (dark lines), we clearly see this transition between an oscillation around almost four initially and around zero finally, whereas the peak-to-peak amplitude of the oscillation remains essentially the same. With increasing excited state decay, the oscillation amplitude decreases because the excited state occupation decays before a complete oscillation around the shifted equilibrium is finished. Once the TLS has returned to its ground state, there is no more coupling to the phonons [see Eq.~\eqref{eq:hamiltonian}]. Thus, in the present model, the oscillations in Fig.~\ref{Fig:4}(b) remain undamped. They would only be damped by taking into account phonon-phonon interactions, i.e., anharmonicities, in the Hamiltonian.

\subsection{Two-pulse excitation}

\subsubsection{Pure dephasing}
\begin{figure}
\includegraphics[width = 1.0\columnwidth]{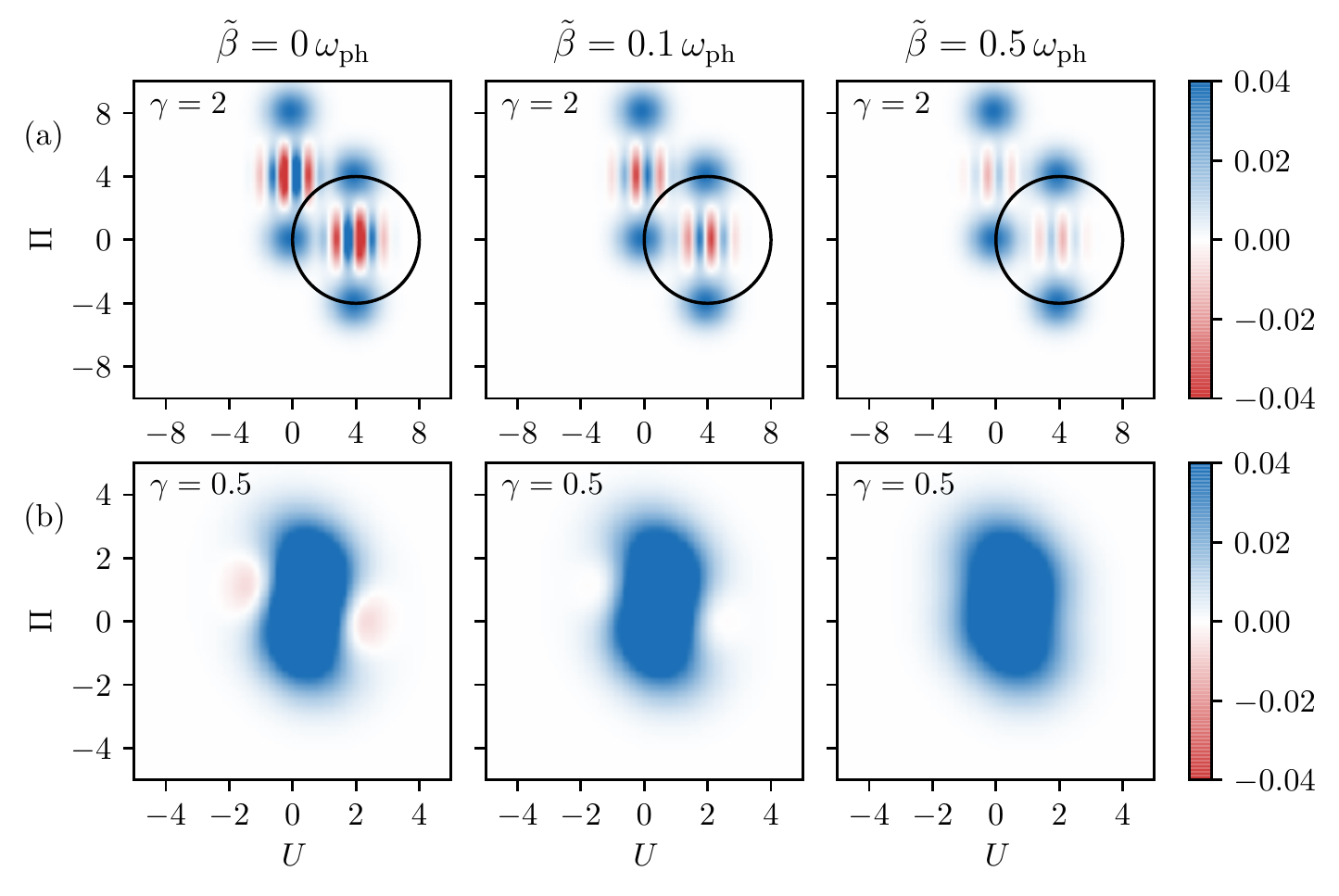}
\caption{Phonon state after a two-pulse excitation. The pulse areas are $\theta_i=\pi/2$, the delay is $t_2-t_1=t_{\rm ph}/2$, and the depicted time is $t/t_{\rm ph}=3/4$ after the second pulse. (a) Wigner function for the coupling strengths (a) $\gamma=2$ and (b) $\gamma=0.5$. The pure dephasing rate increases from $\tilde{\beta}=0$ (left) via $0.1\,\omega_{\rm ph}$ (center) to $0.5\,\omega_{\rm ph}$ (right).}
\label{Fig:5}
\end{figure}

From Eqs.~\eqref{eq:free} and~\eqref{eq:pulse}, we know that the phonon state is only driven by the excited state occupation. Therefore, in a single-pulse excitation the dephasing has no influence on the phonon state. The same holds for phase $\phi$ of the laser pulse, which does not enter in Eq.~\eqref{eq:F_1} after a single pulse. However, in Ref.~\cite{reiter2011gene}, it was shown that by an excitation with two pulses the phonon system can be brought into a statistical mixture of two Schr\"odinger cat states. Cat states are superpositions of coherent states. The interference between these coherent phonon states can be controlled by the relative phase of the laser pulses $\Delta\phi=\phi_1-\phi_2$. To study the influence of a pure dephasing contribution, i.e., a nonvanishing $\beta$, whereas keeping $\Gamma=0$, we consider the same two-pulse excitation as in Ref.~\cite{reiter2011gene} but with $\beta=\tilde{\beta}\neq 0$. Having a look at the schematic picture of the generating functions in Fig.~\ref{Fig:1}(a), we find that the coherence function $Y$ has no direct influence on the phonon state $F$. Therefore, once a phonon state is prepared by a pulse sequence and the population function $C$ is not affected by any decay,  the phonon state behaves periodically with the phonon frequency $\omega_{\rm ph}$, and the dephasing has no impact. But, in the case of a two-pulse excitation, the coherence function $Y$ between the two pulses can act on the phonon state indirectly via $C$. To visualize the influence of a nonvanishing pure dephasing rate on the generation of phonon cat states, in Fig.~\ref{Fig:5}, we plot the Wigner function after a pulse sequence with pulse areas $\theta_1=\theta_2=\pi/2$, delay $t_2-t_1=\pi/\omega_{\rm ph}=t_{\rm ph}/2$, and relative phase $\Delta\phi=\pi/2$. The phonon state is depicted at $t/t_{\rm ph}=3/4$ after the second laser pulse excitation. In Fig.~\ref{Fig:5}(a), we choose a large coupling strength of $\gamma=2$ and increase the pure dephasing rate from $\tilde\beta =0$ (left) via $0.1\,\omega_{\rm ph}$ (center) to $0.5\,\omega_{\rm ph}$ (right). The typical structure of the Schr\"odinger cat states consisting of two Gaussians with a striped structure between them is visible. The striped pattern is oriented along the connection line between the two Gaussians; it takes positive and negative values and is a direct consequence of the quantum-mechanical interference between the two coherent states in the phonon system. We observe two such cat states, one in each electronic subspace (ground and excited states). When increasing the dephasing rate, we find that the interference features faint, which nicely illustrates the loss of coherence during the time interval between the laser pulses. It should be noted that, once a quantum coherence is generated in the phononic system, it is unaffected by the pure dephasing processes acting on the excited state of the TLS. The reason is that the phonon state is only driven by the occupation function $C$, which is not affected by the dephasing. Additionally, the pure dephasing does not alter the coherent phonon states, which still appear as symmetric Gaussians.

In Ref.~\cite{reiter2011gene}, it was also shown that these mixtures of cat states in the phonon system can exhibit squeezing, i.e., the quantum fluctuations of displacement or momentum can fall below the respective vacuum value. This is possible when the coupling strength between the phonon and the TLS is small enough, e.g., $\gamma=0.5$. This effect strongly depends on the relative phase between the two laser pulses $\Delta\phi$ and is a consequence of the quantum interferences. Under the right conditions, the different contributions of the Wigner function overlap in phase space in such a way that the distribution becomes narrower than the vacuum state. Of course, the Heisenberg uncertainty relation has to be fulfilled, therefore, this squeezing of the Wigner function is always accompanied by a broadening in the perpendicular direction. This effect is visible in Fig.~\ref{Fig:5}(b) where we show the same Wigner functions as in Fig.~\ref{Fig:5}(a) but for $\gamma=0.5$. In the case of no pure dephasing $\tilde\beta =0$ (left), the two slightly negative contributions lead to an additional narrowing of the distribution. When increasing the dephasing rate, these negative parts vanish, and the Wigner function becomes entirely positive. It directly follows that the squeezing vanishes because without any interferences, the phonon state is a statistical mixture of coherent states and all quantum effects that could lead to a reduction of the fluctuations disappear.

\begin{figure}
\includegraphics[width = 1.0\columnwidth]{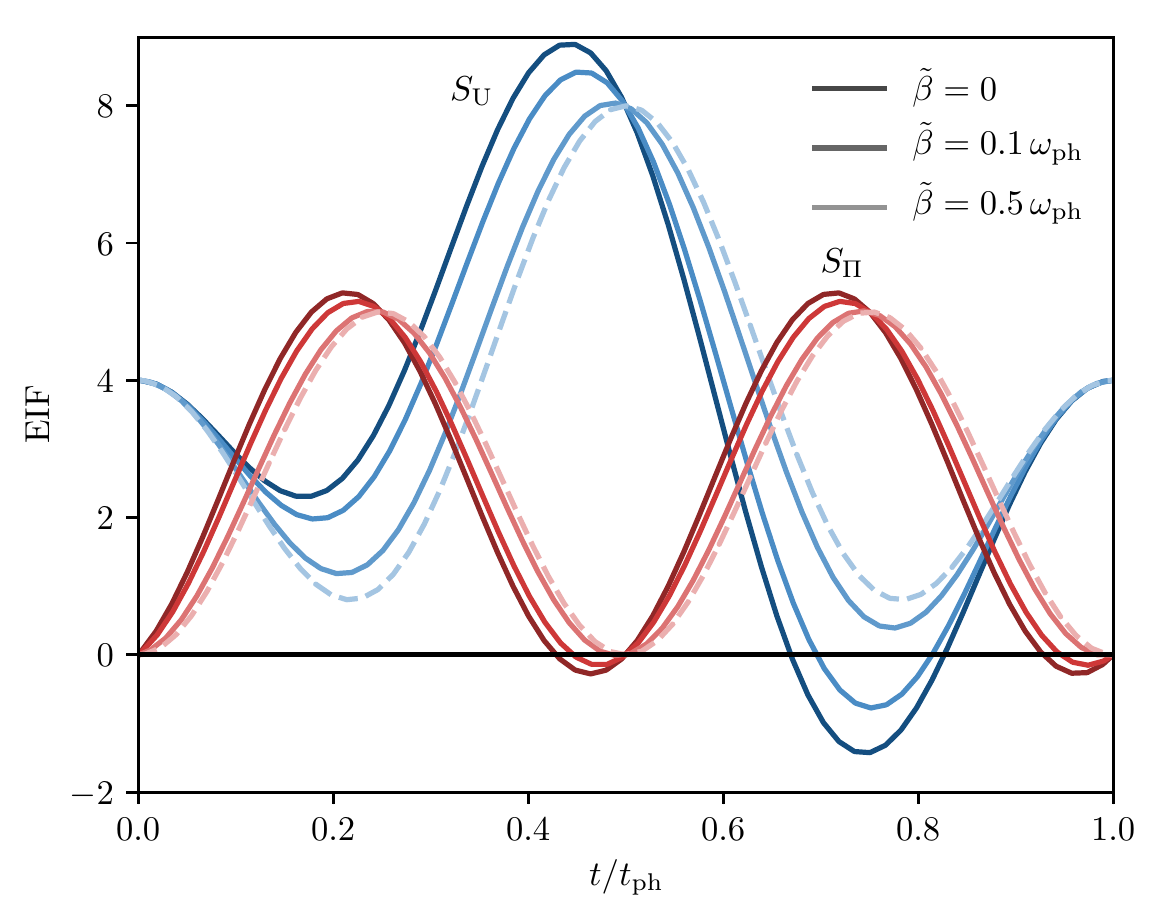}
\caption{EIFs $S_U$ (blue) and $S_\Pi$ (red) as functions of time after the second pulse. The phonon coupling strength is $\gamma=0.5$, and  the pure dephasing rate increases from dark to bright lines. The dashed line is the limiting case of a statistical mixture, i.e., $\tilde{\beta}\gg\omega_{\rm ph}$.}\label{Fig:6}
\end{figure}

To obtain an overview of the influence of pure dephasing on the quantum fluctuations of the phonon states, we consider the excitation-induced fluctuations (EIFs) of the displacement and the momentum defined by~\cite{reiter2011gene,sauer2010latt}
\begin{subequations}
\begin{eqnarray}
S_U &=& \frac{(\Delta u)^2-(\Delta u_{\rm vac})^2}{\gamma^2}\ ,\\
S_\Pi &=& \frac{(\Delta \pi)^2-(\Delta \pi_{\rm vac})^2}{\gamma^2}\ .
\end{eqnarray}\end{subequations}

In Fig.~\ref{Fig:6}, these quantities are plotted as functions of time $t$ after the second pulse. The blue curves show $S_U$, the red ones show $S_\Pi$, and the pure dephasing rate increases from dark to bright colors from $\tilde\beta=0$ via $0.1\,\omega_{\rm ph}$ to $0.5\,\omega_{\rm ph}$. As already discussed in the context of the Wigner functions, we find that the squeezing shrinks, i.e., the negative values reduce. At the same time, also the maximal positive values of the fluctuations get smaller when increasing the dephasing. This shows that the influence of the quantum interference not only leads to the reduction, but also causes increased fluctuations. This is in direct correspondence with the Heisenberg uncertainty principle. For the largest considered dephasing rate of $\tilde{\beta}=0.5\,\omega_{\rm ph}$ the incoherent limit of a statistical mixture of all four coherent states (dashed lines) is almost reached. The minimum of the momentum's EIF is $S_\Pi=0$ whereas the displacement's EIF remains always positive as is expected for the statistical mixture of four coherent states oscillating around equilibrium positions shifted in the $U$ direction \cite{reiter2011fluc}.

\subsubsection{Influence of excited state decay on cat state dynamics}\label{sec:cat_decay}

The considered generation mechanism of phonon cat states, in general, leads to a state which is partly attributed to the ground state $\left|g\right>$ and partly to the excited state $\left|x\right>$. The phonon state attributed to the ground state oscillates stable in shape because it is not affected by decay and dephasing. Therefore, from the mixture of cat states, the one in the ground state will not change once it is generated. But the state attributed to the excited state will be affected by the decay into the ground state. Although in the considered system it is not possible to generate an ideal cat state only in the excited state, in the following, we will study the evolution of such a state to understand the influence of the excited state decay. In the next section, we will then analyze the full dynamics including the generation of the cat states. 

Mathematically, we can initialize such an ideal cat state by applying two laser pulses with delay $t_2-t_1=t_{\rm ph}/2$, pulse areas $\theta_1=\theta_2=\pi/2$, relative phase $\Delta\phi=\pi/2$, and disregarding decay and dephasing during the phonon state preparation. As already discussed, this excitation scheme will generate one cat state in each electronic subspace. In order to isolate the one that belongs to the excited state, we set $F(\alpha,t=0)=C(\alpha,t=0)$ as initialization after the second pulse. This removes the phonon state associated with the ground state because $C(\alpha)$ carries the entire phonon state attributed to the excited state~\cite{wigger2016quan}. Note that, after this procedure, the phonon state is not normalized any longer. For the dynamics following this initialization, we consider again a nonvanishing decay rate to investigate the transition into the ground state. In this case, the dephasing has again no influence on the phonon dynamics. 

\begin{figure}
\includegraphics[width = 1.0\columnwidth]{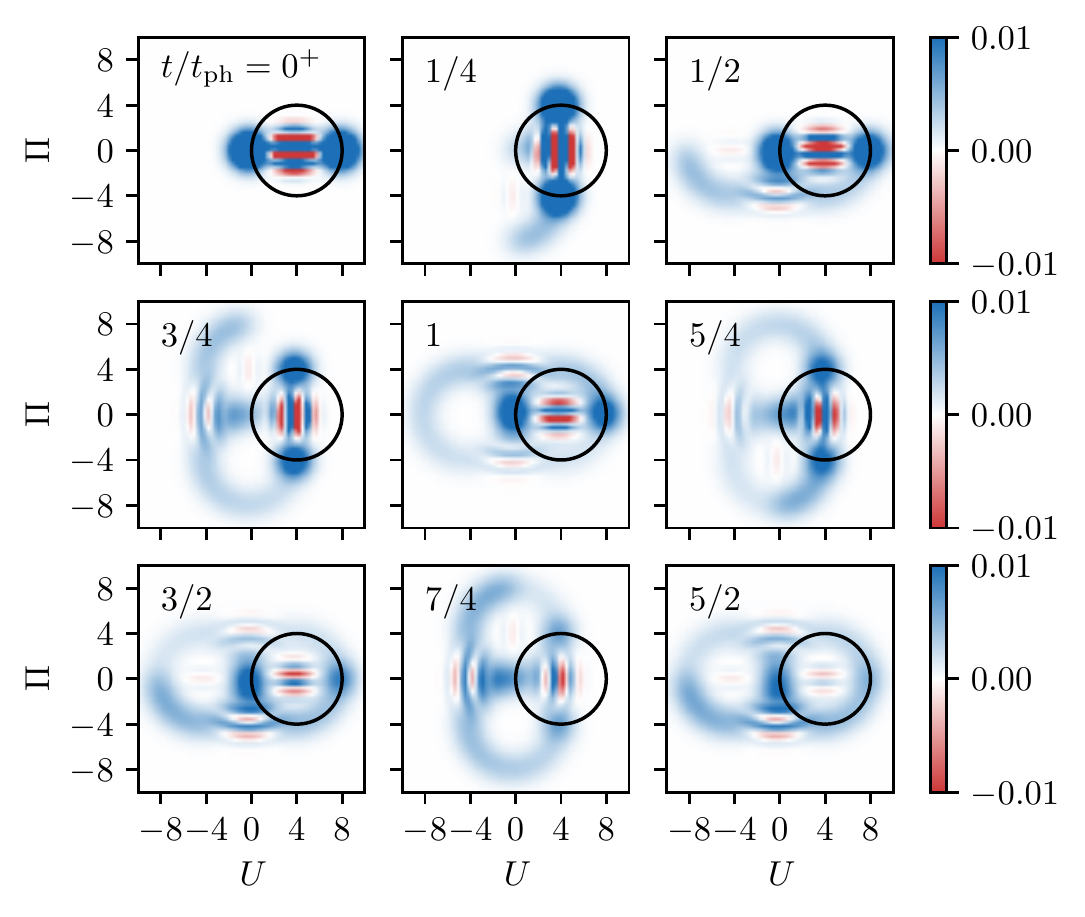}
\caption{Excited state decay induced decay of a cat state. Snapshots of the Wigner function's dynamics for a cat state prepared in the excited state $\left|x\right>$. The decay rate is $\Gamma=0.2\,\omega_{\rm ph}$, and the coupling strength is $\gamma=2$.}
\label{Fig:7}
\end{figure}

Figure~\ref{Fig:7} shows snapshots of the Wigner function's dynamics for an initial cat state in the excited state subspace at $t=0$ and a decay rate of $\Gamma=0.2\,\omega_{\rm ph}$. The natural propagation of the phonon state in the excited state is a rotation around the shifted equilibrium position at $(U,\Pi)=(4,0)$, which is the center of the interference term. The rotating dynamics of the Wigner function is accompanied by a continuous flow into the ground state, i.e., into the unshifted phase space. Because the decay is rather slow compared to the phonon period, the Wigner function evolves into a double-ring structure in the shape of an 8 rotating around its center. For the last depicted time $t/t_{\rm ph} = 5/2$, the system has not yet completely decayed, and we find four interference terms, two at $(0,\pm 4)$, one at the original position of the cat state's interference $(4,0)$, and a very weak one at $(-4,0)$. Already, after the second time step at $t/t_{\rm ph}=1/2$, it can be seen that the interference term of the cat state is transferred together with the coherent contributions into the ground state subsystem as it clearly appears as striped structure around $(0,-4)$. This and the interference on the opposite side of the origin survive the decay process and remain even after the full decay. 

\begin{figure}
\includegraphics[width = 1.0\columnwidth]{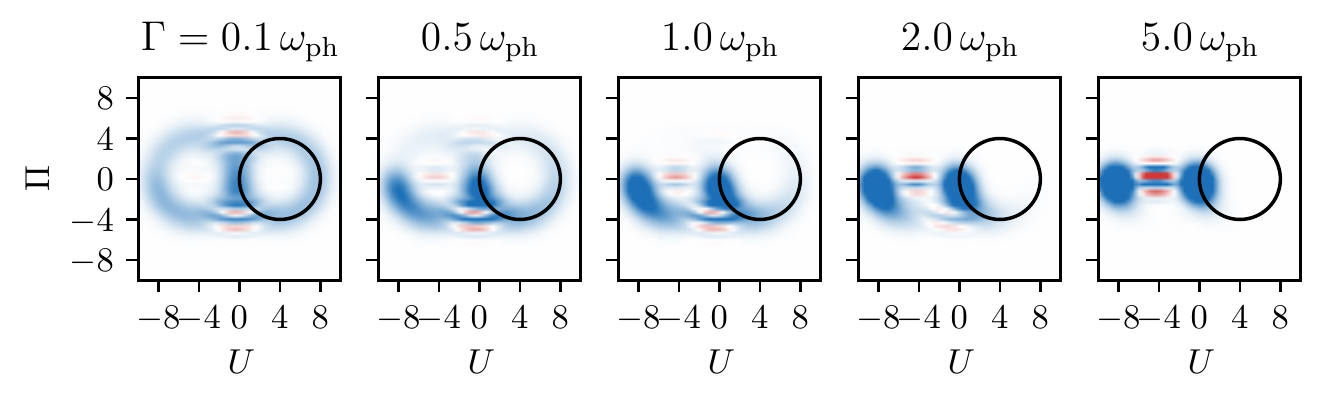}
\caption{Phonon Wigner function resulting from the excited state-induced decay of a cat state at the time $t/t_{\rm ph}=9.5$, i.e., after the excited state occupation has completely decayed for an electron-phonon coupling $\gamma=2$ and five different values of the decay constant $\Gamma$.} \label{Fig:8}
\end{figure}

Figure~\ref{Fig:8} shows the Wigner function at the time $t/t_{\rm ph}=9.5$, i.e., at a time when the excited state occupation has completely decayed, for different values of the decay constant $\Gamma$. Such as in the case of the single coherent state [see Fig.~\ref{Fig:4}], we observe a transition from a ringlike structure for a decay time much longer than the oscillation period to a more localized structure for very short decay times. Here, the ringlike structure consists of two circles attached to each other. For small values of the decay constant, we observe interference patterns in the regions around  $(U,\Pi)=(0,\pm 4)$. When increasing the decay constant, the distribution of the Wigner function along the circles becomes more and more inhomogeneous. Inside the left circle, an interference pattern builds up whereas the patterns between the circles fade away, first in the region of positive $\Pi$ and then also for negative $\Pi$. For the very strong decay corresponding to $\Gamma=5\,\omega_{\rm ph}$, the initial cat state in the excited state subspace is almost instantaneously transferred to the ground state subspace by keeping the entire interference between the two peaks. Whereas the initial cat state rotated along the solid circle around the shifted equilibrium position of the excited state subspace, the final cat states rotate around the ground state equilibrium position at the origin.

\begin{figure}
\includegraphics[width = 1.0\columnwidth]{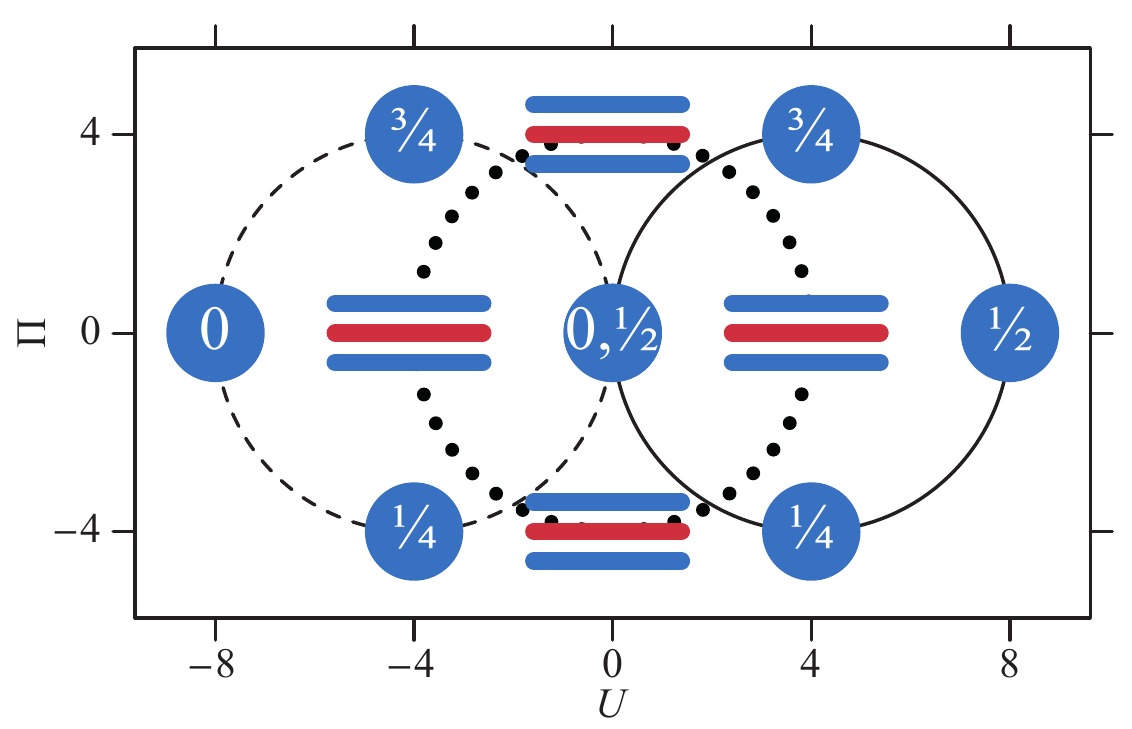}
\caption{Schematic of the Wigner function in the ground state subspace at times $t/ t_{\rm ph}=n+\frac{1}{2}$ ($n\ge1$). The four horizontally aligned cat states are examples for cat states resulting from decay processes at different times. The numbers indicate the time of the decay (in units of $t/t_{\rm ph}$). For a continuous decay, the interference patterns are continuously distributed along the dotted circle.}
\label{Fig:9}
\end{figure}

The previous discussion clearly explains why, in Fig.~\ref{Fig:8}, in the limit of very strong decay, there is an interference pattern around $(U,\Pi)=(-4,0)$, but we still have to understand why this pattern vanishes for weak decay and other patterns at $(U,\Pi)=(0,\pm 4)$ show up. This behavior results from the continuous transfer of the cat state from the excited state subspace to the ground state subspace. So we can imagine the distribution as the sum of many cat states transferred to the ground state at different times. At half-integer times $t/t_{\rm ph}=n+\frac{1}{2}$, the two Gaussians of each of these cat states are aligned horizontally, and they are all separated by $\Delta U=8$, i.e., the diameter of the two circles. Thus, we obtain a superposition of interference patterns which are all aligned horizontally but shifted vertically. This is indicated in Fig.~\ref{Fig:9} where the position at times $t/t_{\rm ph}=n+\frac{1}{2}$ ($n\ge1$) of the cat states resulting from decay processes at four different times $t/t_{\rm ph} = 0, \frac{1}{4}, \frac{1}{2}, \frac{3}{4}$ are plotted schematically together with their respective interference patterns. Since the decay processes occur continuously, the interference patterns are continuously distributed along the dotted circle. Due to the continuous superposition of these vertically shifted interference patterns, inside the solid and dashed circles there is a destructive interference, which removes the interference patterns in these regions. In contrast, at the top and the bottom of the dotted circle, there is a stationary phase of the interference pattern due to the horizontal slope of the circles resulting in constructive interference. This explains why, in the case of weak decay, when the distribution of the Wigner function along the circles is almost homogeneous, interference patterns remain around  $(U,\Pi)=(0,\pm 4)$, whereas they vanish around $(U,\Pi)=(\pm 4,0)$ as it was produced at a later time. With increasing the decay constant, the decay at later times is continuously reduced, which first leads to the vanishing of the interference pattern around $(U,\Pi)=(0,4)$ and finally around $(U,\Pi)=(0,-4)$, whereas the pattern around $(U,\Pi)=(-4,0)$ remains because of the negligible vertical shift of the contributing cat states. 

\subsubsection{Cat state generation with excited state decay and dephasing}

Finally we go a step further and study the influence of a nonvanishing decay rate $\Gamma$ also during the generation process of the phononic Schr\"odinger cat states. We do not consider an additional pure dephasing as the decay already has a dephasing influence on the coherence function with $\beta=\Gamma/2$. With what we have learned so far, we expect a more complex structure of the Wigner functions of the phonon states after the second laser pulse because the single contributions will not be coherent any more and the characteristic tail structures from Sec.~\ref{sec:decay} will appear.

\begin{figure}
\includegraphics[width = 1.0\columnwidth]{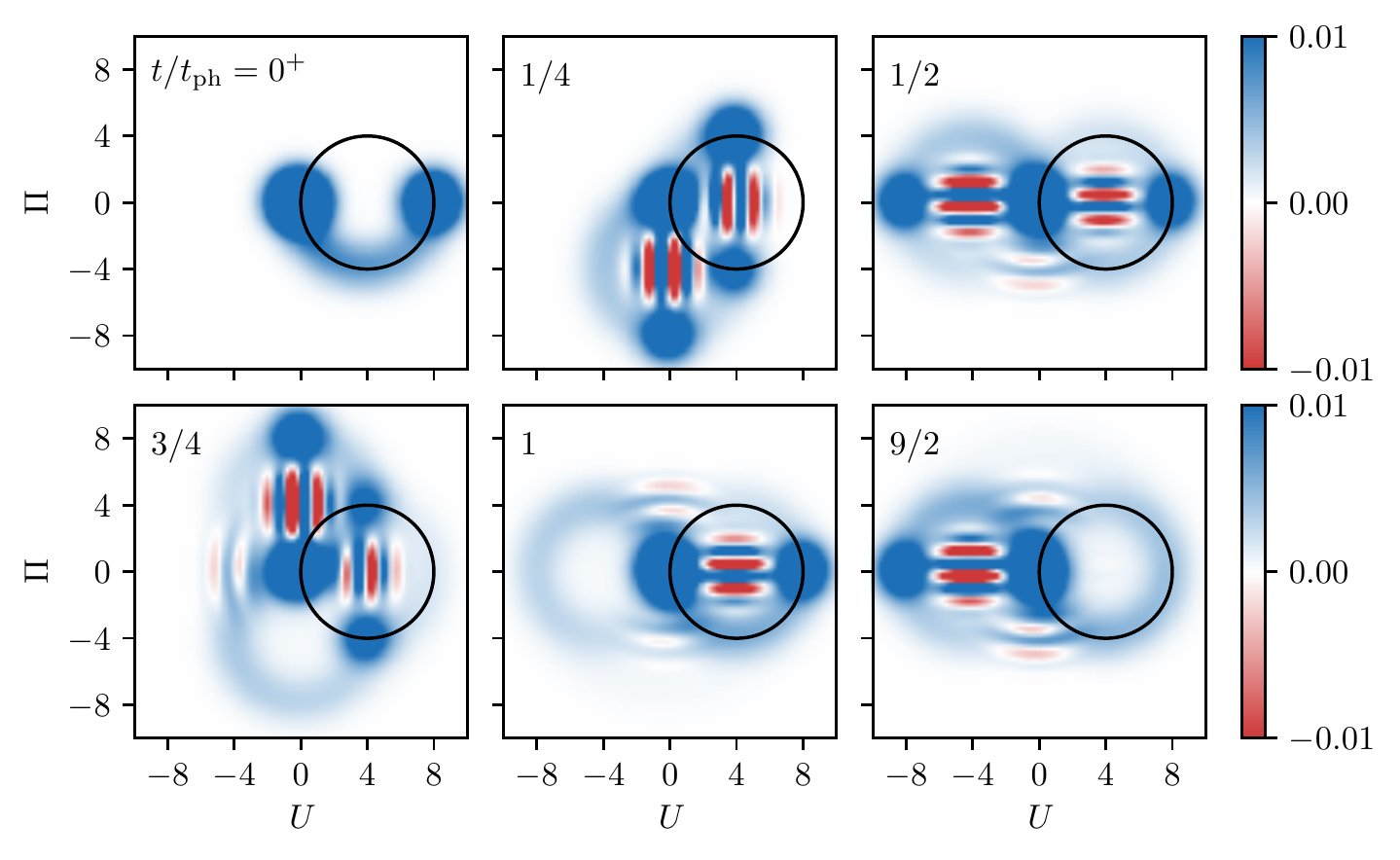}
\caption{Cat state generation with excited state decay and dephasing. Snapshots of the Wigner function's dynamics during the first phonon period after the second laser pulse ($t=0\,\to\,t=t_{\rm ph}$) and at the time $t/t_{\rm ph}=9/2$ (bottom right). The decay rate is $\Gamma=0.2\,\omega_{\rm ph}$, and the coupling strength is $\gamma=2$.}
\label{Fig:10}
\end{figure}

For the dynamics of the Wigner function shown in Fig.~\ref{Fig:10}, we consider a decay rate of $\Gamma=0.2\,\omega_{\rm ph}$. We see that, after the second laser pulse at $t=0$, the Gaussians are accompanied by a half circle distribution as a result of the decay between the two laser pulses. Despite the nonvanishing decay rate, still a major part of the phonon state is coherent, and interferences between these states build up in the following dynamics. As discussed in Sec.~\ref{sec:cat_decay}, the phonon part attributed to the ground state is not affected by the decay, whereas the part in the excited state decays as previously explained, i.e., it evolves into the eight-shaped structure. Compared to the shape-invariant cat state in the ground state subspace, the overall amplitude of this part of the Wigner function is rather small because the distribution covers a larger area in phase space. So the main contribution of the phonon state after the full decay of the excited state is a single cat state as depicted in Fig.~\ref{Fig:10} at $t/t_{\rm ph}=9/2$.

It is now possible to bring this state into the excited state potential by applying a third laser pulse with pulse area of $\theta_3=\pi$, which entirely inverts the TLS. When this is performed at full phonon periods after the second pulse, i.e., at $t/t_{\rm ph}=n$ ($n > 4$), it leads to approximately the situation artificially prepared in Fig.~\ref{Fig:7} at $t=0$. After the following decay into the ground state, the entire phonon state will have evolved into the eight-shaped distribution depicted in Fig.~\ref{Fig:8} at the end of the relaxation. Any other choice of the time or pulse area for the third exciting laser pulse will lead to a more complex combination for the phonon cat state in the excited state, meaning that only one of the two coherent states will move on the circle depicted in the figures. The other one will move on a larger circle around the same center.

\section{Conclusions}
We have studied an optically driven single-photon emitter coupled to a single discrete phonon mode. Under the assumption that the decay and dephasing time scale of the TLS is in the range of the phonon period, we have systematically analyzed the preparation of phonon quantum states by ultrafast optical excitations of the TLS. After a single-pulse excitation, only the decay rate has an impact on the phonon state. During the decay of the excited state, the Wigner function of the phonon state transforms from a coherent Gaussian into a ring-shaped structure, which reduces the amplitude of the lattice displacement. For a two-pulse excitation, also the dephasing between the two pulses acts on the phonon state. For strong dephasing rates, the phonon system loses its coherence, and the interferences representing the Schr\"odinger cat states vanish. This also removes the possibility of the phonon state to exhibit squeezing. When taking into account decay and dephasing during a two-pulse excitation, more complex Wigner functions emerge, and we have shown that interferences in the phonon system can be transferred from the excited state subspace of the TLS to its ground state subspace. Because of destructive interference of cat states transferred to the ground state subspace at different times, the interference patterns may be blurred in some regions of phase space. At the same time they survive in other regions. Our paper shows that there is a subtle interplay amng different time scales, in particular, the phonon oscillation period, the decay and dephasing times, and the delay of the exciting laser pulses, which can be used to tailor the final phonon quantum state.


\section*{Appendix}

The recursion relations connecting the values of the characteristic functions $\overline{Y}$, $\overline{C}$, and $\overline{F}$ immediately before and after the  pulse $j$, i.e., from $t_j^-$ to $t_j^+$, read
\begin{widetext}
\begin{subequations}
\begin{eqnarray}
\overline{Y}_j^+(\alpha) &=&  
\frac{1}{2} [ 1 + \cos(\theta_j) ] \overline{Y}_{j}^{-}(\alpha) \notag\\
&& + \sin^2 \left(\frac{\theta_j}{2} \right)\overline{Y}_{j}^{-\ast}\left(2\gamma e^{-i\omega_{\rm ph} t_j}-\alpha\right) \exp\left\{ i2\phi_j + 2\gamma{\rm Re} \left[ \left(\alpha e^{i \omega_{\rm ph} t}- \gamma\right) \right] \right\} \notag \\
&&+ \left[ \overline{F}_{j}^{-}\left(\alpha - \gamma e^{-i\omega_{\rm ph} t_j}\right) 
	- 2 \overline{C}_{j}^{-}\left(\alpha - \gamma e^{-i\omega_{\rm ph} t_j} \right)e^{-i2\gamma{\rm Im}\left( \alpha e^{i \omega_{\rm ph} t_j}\right)} \right] 
\frac{i}{2} \sin\left(\theta_j\right) \exp\left(i\phi_j + \gamma \alpha e^{i \omega_{\rm ph} t_j}\right) \\
\overline{C}_j^+ (\alpha) &=& 
\overline{C}_{j}^{-}(\alpha) \notag \\
&& + \exp\left[ i 2 \gamma {\rm Im}\left(\alpha e^{i \omega_{\rm ph} t_j}\right)\right]
 \Bigg( \sin^2 \left(\frac{\theta_j}{2}\right) 
 \left\{ \overline{F}_{j}^{-}(\alpha) 
 		- 2\overline{C}_{j}^{-}(\alpha)\exp\left[ -i2 \gamma{\rm Im}\left( \alpha e^{i \omega_{\rm ph} t_j}\right) \right] \right\} \notag \\
&& \hspace{4cm} - \frac{i}{2} \sin\left( \theta_j \right) 
\bigg\{ \overline{Y}_{j}^{-} \left( \alpha + \gamma e^{-i \omega_{\rm ph} t_j}\right) 
\exp \left[-i\phi_j - \left(\gamma^2+\gamma \alpha e^{i \omega_{\rm ph} t_j}\right) \right] \notag \\
&&\hspace{5.8cm} -\overline{Y}_{j}^{-\ast} \left(-\alpha + \gamma e^{-i\omega_{\rm ph} t_j}\right)
\exp \left[i\phi_j - \left(\gamma^2- \gamma \alpha^* e^{-i\omega_{\rm ph} t_j}\right) \right] \bigg\} \Bigg) \mbox{\quad}\\
\overline{F}_j^+ (\alpha) &=& \overline{F}_j^-(\alpha)\,.
\end{eqnarray}
\label{eq:pulse}
\end{subequations}
\end{widetext}


%

\end{document}